\documentclass[12pt]{article}
\usepackage{bbm,latexsym}
\newcommand{\be}{\begin{equation}}
\newcommand{\ee}{\end{equation}}
\newcommand{\ba}{\begin{eqnarray}}
\newcommand{\ea}{\end{eqnarray}}
\newcommand{\no}{\nonumber\\}
\newcommand{\lesssim}{\:\mbox{\raisebox{-3pt}{$\stackrel%
{\displaystyle <}{\sim}$}}\:}

\newcommand{\mbb}{m_{\beta\beta}}
\newcommand{\dsm}{\Delta m^2_\odot}
\newcommand{\dam}{\Delta m^2_\mathrm{atm}}
\newcommand{\mm}{{\bar m}^2}
\begin{document}
\title{\normalsize \hfill UWThPh-2004-31 \\[1cm] \LARGE
Softly broken lepton number $L_e - L_\mu - L_\tau$ \\
with non-maximal solar neutrino mixing}
\author{Walter Grimus\thanks{E-mail: walter.grimus@univie.ac.at} \\
\setcounter{footnote}{6}
\small Institut f\"ur Theoretische Physik, Universit\"at Wien \\
\small Boltzmanngasse 5, A--1090 Wien, Austria \\*[3.6mm]
Lu\'\i s Lavoura\thanks{E-mail: balio@cfif.ist.utl.pt} \\
\small Universidade T\'ecnica de Lisboa
and Centro de F\'\i sica Te\'orica de Part\'\i culas \\
\small Instituto Superior T\'ecnico,
P--1049-001 Lisboa, Portugal \\*[4.6mm] }

\date{20 October 2004}

\maketitle

\begin{abstract}
We consider the most general neutrino mass matrix
which leads to $\theta_{13} = 0$,
and present the formulae needed
for obtaining the neutrino masses and mixing parameters in that case.
We apply this formalism to a model
based on the lepton number $\bar L = L_e - L_\mu - L_\tau$
and on the seesaw mechanism.
This model needs only one Higgs doublet
and has only two right-handed neutrino singlets.
Soft $\bar L$ breaking is accomplished by the Majorana mass terms
of the right-handed neutrinos;
if the $\bar L$-conserving and $\bar L$-breaking mass terms
are of the same order of magnitude,
then it is possible to obtain a consistent $\bar L$ model
with a solar mixing angle significantly smaller than $45^\circ$.
We show that the predictions of this model,
$m_3 = 0$ and $\theta_{13} = 0$,
are invariant under the renormalization-group running
of the neutrino mass matrix.
\end{abstract}

\newpage

\section{Introduction}

In recent times there has been enormous experimental progress
in our knowledge of the mass-squared differences
and of the mixing of light neutrinos---for a review see,
for instance,
\cite{maltoni}.
Unfortunately,
to this progress on the experimental
and phenomenological---i.e.\ neutrino oscillations \cite{pontecorvo}
and the MSW effect \cite{MSW}---fronts
there has hardly been a counterpart in our theoretical understanding
of neutrino masses and lepton mixing---for a review see,
for instance,
\cite{altarelli}.

It has been conclusively shown
that the lepton mixing matrix is substantially different
from the quark mixing matrix.
Whereas the solar mixing angle,
$\theta_{12}$,
is large---$\theta_{12} \sim 33^\circ$---and the atmospheric mixing angle,
$\theta_{23}$,
could even be maximal,
the third mixing angle,
$\theta_{13}$,
is small---there is only an upper bound on it which,
according to~\cite{maltoni},
is given at the 3$\sigma$ level by $\sin^2 \theta_{13} < 0.047$.
The true magnitude of $\theta_{13}$
will be crucial in the future experimental exploration of lepton mixing,
and it is also important for our theoretical understanding
of that mixing---see,
for instance,
\cite{barr}.

In this letter we contemplate the possibility
that at some energy scale a flavour symmetry exists
such that $\theta_{13}$ is exactly zero.
It has been shown~\cite{GJKLST} that,
in the basis where the charged-lepton mass matrix is diagonal,
the most general neutrino mass matrix
which yields $\theta_{13} = 0$ is given,
apart from a trivial phase convention \cite{GJKLST},
by
\be
\label{f1}
\mathcal{M}_\nu = \left( \begin{array}{ccc}
X &
\sqrt{2} A \cos{\left( \gamma / 2 \right)} &
\sqrt{2} A \sin{\left( \gamma / 2 \right)} \\
\sqrt{2} A \cos{\left( \gamma / 2 \right)} &
B + C \cos{\gamma} &
C \sin{\gamma} \\
\sqrt{2} A \sin{\left( \gamma / 2 \right)} &
C \sin{\gamma} &
B - C \cos{\gamma}
\end{array} \right),
\ee
with parameters $X$,
$A$,
$B$,
and $C$ which are in general complex.
The mass matrix~(\ref{f1}),
but not necessarily the full Lagrangian,
enjoys a $\mathbbm{Z}_2$ symmetry~\cite{GJKLST,softD4}---see
also~\cite{ustron01}---defined by
\be
S \left( \gamma \right) \mathcal{M}_\nu\,
S \left( \gamma \right) = \mathcal{M}_\nu,
\label{sym}
\ee
with an orthogonal $3 \times 3$ matrix
\be
S \left( \gamma \right) = \left( \begin{array}{ccc}
1 & 0 & 0 \\
0 & \cos{\gamma} & \sin{\gamma} \\
0 & \sin{\gamma} & - \cos{\gamma}
\end{array} \right)
\ee
which satisfies
$S \left( \gamma \right) = \left[ S \left( \gamma \right) \right]^T$
and $\left[ S \left( \gamma \right) \right]^2 = \mathbbm{1}$.
We may remove two unphysical phases from~(\ref{f1}),
e.g.\ by choosing $X$ and $A$ to be real,
and then there are seven real parameters in that mass matrix.
Those seven parameters correspond
to the following seven physical quantities:
the three neutrino masses $m_{1,2,3}$,
the solar and atmospheric mixing angles,
and two Majorana phases.
The only prediction of the mass matrix~(\ref{f1}) is $\theta_{13} = 0$;
however,
that prediction entails the non-observability of the Dirac phase
in lepton mixing---in general there are \emph{nine} physical quantities
in the neutrino masses and mixings.

Expressed in terms of the parameters\footnote{The procedure
for obtaining the neutrino masses and the lepton mixing matrix
from the parameters of a fully general neutrino mass matrix
has been given in~\cite{aizawa}.} of $\mathcal{M}_\nu$,
one obtains masses given by~\cite{GJKLST,GLZ2}
\be
\label{m3}
m_3 = \left| B - C \right|
\ee
and
\be
\label{masses}
m_{1,2}^2 = \frac{1}{2} \left[ |X|^2 + |D|^2 + 4 |A|^2 \mp 
\sqrt{\left( |X|^2 + |D|^2 + 4 |A|^2 \right)^2 -
4 \left| XD - 2 A^2 \right|^2}\right],
\ee
with
\be
D \equiv B + C;
\ee
while the mixing angles are given by
\be
\theta_{23} = \left| \gamma/2 \right|
\ee
and
\ba
\left( m_2^2 - m_1^2 \right) \sin{2 \theta_{12}} &=&
2 \sqrt{2} \left| X^\ast A + A^\ast D \right|,
\label{sinsol} \\
\left( m_2^2 - m_1^2 \right) \cos{2 \theta_{12}} &=& |D|^2 - |X|^2.
\label{cossol}
\ea
The only Majorana phase which---for $\theta_{13} = 0$---plays a role
in neutrinoless $\beta\beta$ decay 
is the phase $\Delta =
\arg{\left[ \left( U_{e2} / U_{e1} \right)^2 \right]}$,
where $U_{e1}$ and $U_{e2}$ are matrix elements
of the lepton mixing
(PMNS~\cite{pontecorvo,mns})
matrix $U$.
The phase $\Delta$ is given by~\cite{leptogenesis}
\be\label{Delta}
8\, \mbox{Im} \left( X^\ast D^\ast A^2 \right) =
m_1 m_2 \left( m_1^2 - m_2^2 \right) \sin^2{2 \theta_{12}} \sin{\Delta}.
\ee
The other Majorana phase is practically 
unobservable~\cite{rodejohann1}.\footnote{When $|X| = |D|$ and 
$X^\ast A = - A^\ast D$ one should use, 
instead of~(\ref{masses}) and~(\ref{sinsol})--(\ref{Delta}),
$m_1 = m_2 = \sqrt{|X|^2 + 2\, |A|^2}$,
$\Delta = \pi$,
and $\cos{2 \theta_{12}} = |X| / m_1$.}

In specific models with $\theta_{13} = 0$,
the neutrino mass matrix~(\ref{f1}) is further restricted:
\begin{itemize}
\item The $\mathbbm{Z}_2$ model of~\cite{GLZ2},
which is based on the non-Abelian group $O(2)$~\cite{su5},
yields maximal atmospheric neutrino mixing, i.e.\ $\gamma = \pi/2$, and
has six physical parameters.
\item
The $D_4$ model of~\cite{GLD4},
which is based on the discrete group $D_4$,
also has $\gamma = \pi/2$ and,
in addition,
it predicts $XC = A^2$.
The number of parameters is four---in that model the Majorana phases
are expressible in terms of the neutrino masses
and of the solar mixing angle.
\item
The softly-broken-$D_4$ model of~\cite{softD4}
is a generalization of the $D_4$ model:
the atmospheric mixing angle is undetermined,
but $XC = A^2$ still holds.
\item
The seesaw model of the first line of Table I of~\cite{low},
which is based on the Abelian group $\mathbbm{Z}_4$,
reproduces matrix~(\ref{f1}) without restrictions. 
\end{itemize}

In this letter we consider the $U(1)$ symmetry
generated by the lepton number
$\bar L \equiv L_e - L_\mu - L_\tau$~\cite{Lbarold}.
It is well known that exact $\bar L$ symmetry
enforces $\theta_{13} = 0$ 
(with $X = B = C = 0$ in~(\ref{f1})),
while an approximate $\bar L$ symmetry tends to produce
either a solar mixing angle too close to $45^\circ$
or a solar mass-squared difference
too close to the atmospheric mass-squared difference~\cite{Lbar}.
A possible way out of this dilemma
is to assume a significant contribution to $U$ from the diagonalization
of the charged-lepton mass matrix~\cite{rodejohann}; another
possibility is a significant
breaking of $\bar L$~\cite{frigerio}.
Here we discuss a $\bar L$ model,
first proposed in~\cite{softLbar},
which makes use of the seesaw mechanism~\cite{seesaw}
with \emph{only two} right-handed neutrino singlets $\nu_R$.
The $U(1)_{\bar L}$ symmetry is softly broken
in the Majorana mass matrix of the $\nu_R$,
but---contrary to what was done
in~\cite{softLbar}---the soft breaking
is assumed here to be rather `strong', in order
to achieve a solar mixing angle
significantly smaller than $45^\circ$.
The model presented in Section~2
predicts a mass matrix~(\ref{f1}) with $B = C$,
i.e.\ it predicts $m_3 = 0$ together with $\theta_{13} = 0$.
We will show in Section~3 that these predictions \emph{are stable}
under the renormalization-group running
from the seesaw scale down to the electroweak scale.
Next,
we show in Section~4 that our model does \emph{not} provide
enough leptogenesis to account for the observed
baryon-to-photon ratio of the Universe.
We end in Section~5 with our conclusions.

\section{The model}

The lepton number $\bar L = L_e - L_\mu - L_\tau$
has a long history in model building~\cite{Lbarold,Lbar}.
In this letter we rediscuss the model of~\cite{softLbar},
which has \emph{only one Higgs doublet},
$\phi$,
and two right-handed neutrinos,
$\nu_{R1}$ and $\nu_{R2}$,
with the following assignments of the quantum number $\bar L$:
\be
\label{Lbar}
\renewcommand{\arraystretch}{1.1}
\begin{array}{c|cccccc}
& \nu_e,\, e & \nu_\mu,\, \mu & \nu_\tau,\, \tau &
\nu_{R1} & \nu_{R2} & \phi \\ \hline
\bar L & 1 & -1 & -1 & 1 & -1 & 0 \end{array}\, .
\ee
This model is a simple extension of the Standard Model
which incorporates the seesaw mechanism~\cite{seesaw}.
The right-handed neutrino singlets have a Majorana mass term 
\be
\mathcal{L}_M = - \frac{1}{2}\, \bar \nu_R M_R C \bar \nu_R^T
+ \mbox{H.c.},
\ee
(where $C$ is the charge-conjugation matrix in spinor space)
with
\be
\label{MR}
M_R = \left( \begin{array}{cc}
R & M \\ M & S \end{array} \right).
\ee
The elements of the matrix $M_R$ are of the heavy seesaw scale.
The entry $M$ in $M_R$ is compatible with $\bar L$ symmetry,
while the entries $R$ and $S$ break that lepton number softly.
The breaking of $\bar L$ is soft
since the Majorana mass terms have dimension three.
Because of the $U(1)$ symmetry associated with $\bar L$,
the neutrino Dirac mass matrix
has the structure~\cite{softLbar}\footnote{We are assuming,
without loss of generality,
the charged-lepton mass matrix to be already diagonal.}
\be
\label{MD}
M_D = \left( \begin{array}{ccc}
a & 0 & 0 \\ 0 & b' & b'' \end{array} \right).
\ee
Then the effective Majorana mass matrix of the light neutrinos
is given by the seesaw formula 
\be
\label{Mnu}
\mathcal{M}_\nu = - M_D^T M_R^{-1} M_D = 
\frac{1}{M^2 - RS} 
\left( \begin{array}{ccc}
S a^2 & - M a b^\prime & - M a b^{\prime\prime} \\
- M a b^\prime & R {b^\prime}^2 & R b^\prime b^{\prime\prime} \\
- M a b^{\prime\prime} & R b^\prime b^{\prime\prime} & R {b^{\prime\prime}}^2
\end{array} \right).
\ee
In the case of $\bar L$ conservation,
i.e.\ when $R = S = 0$,
we have $m_1 = m_2$ and $\theta_{12}$ is $45^\circ$;
this is a well known fact.
Non-zero mass parameters $R$ and $S$
induce $\Delta m^2_\odot \equiv m_2^2 - m_1^2 \neq 0$
and allow a non-maximal solar mixing angle.\footnote{The case
of non-zero $R$ and $S$ has also been considered in~\cite{goh}.} 
The phases of $a$,
$b^\prime$,
and $b^{\prime\prime}$ are unphysical;
in the following we shall assume those parameters
to be real and positive.
The only physical phase is~\cite{softLbar}
\be
\label{alpha}
\alpha \equiv \arg \left( R^\ast S^\ast M^2 \right).
\ee
$CP$ conservation is equivalent to $\alpha$ being a multiple of $\pi$.
Defining $d \equiv M^2 - R S$ and
$b \equiv \sqrt{{b^\prime}^2 + {b^{\prime\prime}}^2}$,
we see that~(\ref{Mnu}) has the form~(\ref{f1}) with
\be
\label{XAB}
X = \frac{S a^2}{d}\, , \quad 
A = - \frac{M a b}{\sqrt{2} d}\, , \quad
B = C = \frac{R b^2}{2 d}\, ,
\ee
and
\be
\cos{\frac{\gamma}{2}} = \frac{b^\prime}{b}\, , \quad
\sin{\frac{\gamma}{2}} = \frac{b^{\prime\prime}}{b}\, .
\ee
As advertised in the introduction,
$C = B$ and therefore $m_3 = 0$, 
while $X$,
$A$,
and $B$ are independent parameters.
Since $m_3 = 0$ the neutrino mass spectrum displays inverted hierarchy. 
The present model has five real parameters---$|X|$,
$|A|$, 
$|B|$, 
$\gamma$,
and $\alpha$---which correspond to the physical observables $m_{1,2}$,
$\theta_{12}$,
$\theta_{23}$,
and the Majorana phase $\Delta$
(the second Majorana phase is unphysical in this case because $m_3 = 0$).

Let us now perform a consistency check
by using all the available input from the neutrino sector. 
We have the following observables at our disposal:
the effective Majorana mass in neutrinoless $\beta\beta$ decay $\mbb$,
the solar mass-squared difference $\dsm$,
the atmospheric mass-squared difference $\dam$,
the solar mixing angle $\theta_{12}$,
and the atmospheric mixing angle $\theta_{23}$. 
In a three-neutrino scenario the definition of $\dam$ is not unique;
we define $\mm \equiv \left( m_1^2 + m_2^2 \right) / 2$
and use $\dam \simeq \mm$,
which is valid in this model because of the inverted mass hierarchy. 
The relation $\gamma = 2 \theta_{23}$
plays no role in the following discussion,
which consists in determining the four parameters $|X|$,
$|A|$,
$|D| = 2\,|B|$,
and $\alpha$ as functions of the four observables $\mbb$,
$\dsm$,
$\dam$,
and $\theta_{12}$.
We note that,
because of~(\ref{alpha}) and~(\ref{XAB}),
$\alpha = \arg{\left( D^\ast X^\ast A^2 \right)}$.

We first note that,
in~(\ref{f1}),
$\mbb$ is just given by $|X|$:
\be
|X| = \mbb.
\label{mbb}
\ee
We then use~(\ref{cossol}) to write
\be
\label{D} 
|D|^2 = \mbb^2 + \dsm \cos{2 \theta_{12}}.
\ee
From~(\ref{masses}),
$\bar m^2 = 2 |A|^2 + \left( |X|^2 + |D|^2 \right) / 2$,
hence 
\be
\label{A}
|A|^2 = \frac{1}{2} \left( \mm - \mbb^2 - 
\frac{1}{2}\, \dsm \cos{2 \theta_{12}} \right).
\ee
Since $|A| \geq 0$ we have the bound
\be
\label{bound}
\mbb^2 \leq \mm - \frac{1}{2}\, \dsm \cos{2 \theta_{12}}.
\ee
In order to find $\alpha$ we start from~(\ref{sinsol}),
writing
\be
\left( \dsm \right)^2 \sin^2{2 \theta_{12}}
= 8\, |A|^2 \left( |X|^2 + |D|^2 + 2\, |X|\, |D|\, \cos{\alpha} \right).
\ee
We define
\ba
p &\equiv& \frac{2 \mbb^2}{\dsm \cos{2 \theta_{12}}}, \\
\rho &\equiv& \frac{2 \mm}{\dsm \cos{2 \theta_{12}}},
\ea
and obtain
\be
\label{cosalpha}
\cos \alpha = \left[ - \left( 1 + p \right)
+ \frac{\tan^2{2 \theta_{12}}}{2 \left( \rho - 1 - p \right)} \right]
\frac{1}{\sqrt{p \left( p + 2 \right)}}.
\ee
Thus we have expressed all the parameters of the model
in terms of physical quantities.

The parameter $\rho$ is known and it is quite large:
using the mean values of the mixing parameters~\cite{maltoni}
$\dsm \sim 8.1 \times 10^{-5}\, \mbox{eV}^2$, 
$\dam \sim 2.2 \times 10^{-3}\, \mbox{eV}^2$, 
and $\sin^2{\theta_{12}} \sim 0.30$,
we find $\rho \sim 136$.
The parameter $p$,
on the other hand,
is unknown.
Equation~(\ref{cosalpha}) requires that
a non-zero range $\left[ p_-,\, p_+ \right]$ for $p$ exists
for which the right-hand side of that equation lies
in between $-1$ and $+1$.
One finds that
\be
p_\pm = \frac{\rho}{2} \left( 1 + \cos^2{2 \theta_{12}}
\pm \sin^2{2 \theta_{12}}\, \sqrt{1 - \frac{1}{\rho^2 \cos^2{2 \theta_{12}}}}
\,\right) - 1.
\ee
Since $\rho$ is large this can be approximated by
\be
p_- \simeq \rho \cos^2{2 \theta_{12}} - 1, \quad
p_+ \simeq \rho - 1,
\ee
or
\be
\label{bounds}
\mm \cos^2{2 \theta_{12}} - \frac{\dsm \cos{2 \theta_{12}}}{2}
\, \lesssim \, \mbb^2 \, \lesssim \,
\mm - \frac{\dsm \cos{2 \theta_{12}}}{2}.
\ee
In this approximation,
the upper bound on $\mbb$ coincides with the one in~(\ref{bound}).
With good accuracy we have in this model
\be
\label{mbbbound}
\sqrt{\dam} \cos{2 \theta_{12}} \, \lesssim \, \mbb \, \lesssim \, \sqrt{\dam}.
\ee
This is one of the predictions of the model. 
Thus,
if the claim $\mbb > 0.1\, \mbox{eV}$ of~\cite{klapdor} is confirmed,
then the present model will be ruled out
since $\sqrt{\dam} \sim 0.047\, \mbox{eV}$. 

From~(\ref{sinsol}),
(\ref{cossol}),
and (\ref{XAB}) we find
the following expression for the solar mixing angle: 
\be
\label{solarmix}
\tan{2 \theta_{12}} = \frac{2 \left| M \right| a b}
{\left| R \right| b^2 - \left| S \right| a^2}\, 
\frac{\left| \left| R \right| b^2 + \left| S \right| a^2 e^{i\alpha} \right|}
{\left| R \right| b^2 + \left| S \right| a^2}.
\ee
This equation shows that non-maximal solar neutrino mixing
is easily achievable
when $|R|$ and $|S|$ are of the same order of magnitude as $|M|$.
This is what we mean with `strong' soft $\bar L$ breaking,
namely that the Majorana mass terms which violate $\bar L$ softly
(i.e.\ $R$ and $S$)
and the one which conserves $\bar L$
(i.e.\ $M$)
are of the same order of magnitude.\footnote{In~\cite{softLbar}
we assumed $|R|,\, |S| \ll |M|$
and ended up with almost-maximal solar mixing,
which was still allowed by the data at that time.}

One may ask whether it is possible to evade this feature
and assume $|R|,\, |S| \ll |M|$.
In that case,
since experimentally $\tan{2 \theta_{12}} \simeq 2.3$,
and since the second fraction in the right-hand side of~(\ref{solarmix})
cannot be larger than 1,
we would conclude that $b / a\, \sim |M / R|$.
But then $|R| b^2$ would be much larger than $|S| a^2$
and therefore $|D| \gg |X|$ which,
from~(\ref{mbb}) and~(\ref{D}),
means that $\mbb^2 \ll \dsm \cos{2 \theta_{12}}$.
This contradicts our previous finding that $\mbb^2$
must be of the order of magnitude of $\dam$.
We thus conclude that the hypothesis $|R|,\, |S| \ll |M|$
is incompatible with the experimental data.

\section{Radiative corrections}

We have not yet taken into account the fact
that the energy scale where $\bar L$-invariance holds
and the mass matrices $M_D$ and $\mathcal{M}_\nu$
have the forms~(\ref{MD}) and~(\ref{Mnu}),
respectively,
is the seesaw scale.
Since our model has \emph{only one} Higgs doublet,
the relation between
the mass matrix $\mathcal{M}_\nu^{(0)}$ at the seesaw scale
and the mass matrix $\mathcal{M}_\nu$ at the electroweak scale
is simply given by~\cite{chankowski}
\be
\mathcal{M}_\nu = I \mathcal{M}_\nu^{(0)} I,
\ee
where $I$ is a diagonal, positive, and non-singular matrix,
since the charged-lepton mass matrix is diagonal.
Now,
suppose there is a vector $u^{(0)}$
such that $\mathcal{M}_\nu^{(0)} u^{(0)} = 0$.
Then the vector $u \equiv I^{-1} u^{(0)}$
is an eigenvector to $\mathcal{M}_\nu$
with eigenvalue zero.\footnote{This statement would still be true
for a non-diagonal matrix $I$.} Moreover,
if one entry of $u^{(0)}$ is zero,
then the corresponding entry of $u$ is zero as well,
due to $I$ being diagonal.
We stress that these observations
only hold for eigenvectors with eigenvalue zero.

Applying this to the present model,
we find that $m_3 = 0$ together with $\theta_{13} = 0$
are predictions \emph{stable under the renormalization-group evolution}.
The matrices $\mathcal{M}_\nu$ and $\mathcal{M}_\nu^{(0)}$
are related through $M_D = M_D^{(0)} I$,
where $M_D^{(0)}$ is the neutrino Dirac mass matrix at the seesaw scale;
again,
due to $I$ being \emph{diagonal},
both Dirac mass matrices have the same form~(\ref{MD}).
Therefore,
all our discussions in the previous section
hold for the physical quantities at the low (electroweak) scale.

\section{Leptogenesis}

The model in this letter has very few parameters
and only one Higgs doublet.
Therefore,
it allows clear-cut predictions
for leptogenesis---for reviews see,
for instance,
\cite{lepto-reviews}.
It turns out that the computations for this model
resemble closely the ones for the $\mathbbm{Z}_2$ model~\cite{GLZ2},
which were performed in a previous paper~\cite{leptogenesis}.
We give here only the gist of the argument.

Let the matrix $M_R$ in~(\ref{MR}) be diagonalized
by the $2 \times 2$ unitary matrix
\be
V = \left( \begin{array}{cc}
c^\prime e^{i \omega} & s^\prime e^{i \sigma} \\
- s^\prime e^{i \tau} & c^\prime
e^{i \left( \sigma + \tau - \omega \right)}
\end{array} \right)
\label{V}
\ee
as
\be
V^T M_R V = \mbox{diag} \left( M_1, M_2 \right),
\label{MRdiag}
\ee
with real,
non-negative $M_1$ and $M_2$.
We assume $M_1 \ll M_2$.
In~(\ref{V}),
$c^\prime \equiv \cos{\theta^\prime}$
and $s^\prime \equiv \sin{\theta^\prime}$,
where $\theta^\prime$ is an angle of the first quadrant.
Defining the Hermitian matrix
\be
H \equiv  V^T M_D M_D^\dagger V^\ast,
\ee
the relevant quantity for leptogenesis is~\cite{lepto-reviews}
\be
\mathcal{Q} \equiv \frac{\mbox{Im} \left[ \left( H_{12} \right)^2 \right]}
{\left( H_{11} \right)^2}.
\ee

One may use as input for leptogenesis
the heavy-neutrino masses $M_{1,2}$ together with $m_{1,2}$,
$\theta_{12}$,
and the Majorana phase $\Delta$.
One can demonstrate that $a$
and $b = \sqrt{{b^\prime}^2 + {b^{\prime\prime}}^2}$ satisfy
\be
a^2 b^2 = m_1 m_2 M_1 M_2
\label{product}
\ee
and
\ba
\lefteqn{
\left| s_{12}^2 m_1 + c_{12}^2 m_2 e^{i \Delta} \right|^2 a^4 
+ \left| c_{12}^2 m_1 + s_{12}^2 m_2 e^{i \Delta} \right|^2 b^4}
\no && 
= m_1^2 m_2^2 \left( M_1^2 + M_2^2 \right)
- 2 m_1 m_2 M_1 M_2 c_{12}^2 s_{12}^2
\left| m_1 - m_2 e^{i \Delta} \right|^2,
\label{sum}
\ea
where $c_{12} \equiv \cos{\theta_{12}}$
and $s_{12} \equiv \sin{\theta_{12}}$.
By using~(\ref{product}) and~(\ref{sum}),
one finds the values of $a$ and $b$ from the input,
with a twofold ambiguity only.
Then $\theta^\prime$ is given by
\be
{c^\prime}^2 - {s^\prime}^2
= \frac{1}{m_1^2 m_2^2 \left( M_1^2 - M_2^2 \right)}
\left( \left| s_{12}^2 m_1 + c_{12}^2 m_2 e^{i \Delta} \right|^2 a^4
- \left| c_{12}^2 m_1 + s_{12}^2 m_2 e^{i \Delta} \right|^2 b^4 \right).
\ee

With $a$ and $b$ known,
$\mathcal{Q}$ is found as a function of the input by use of
\ba
H_{11} &=& a^2 {c^\prime}^2 + b^2 {s^\prime}^2,
\\
\mbox{Im} \left[ \left( H_{12} \right)^2 \right] &=&
\left( b^2 - a^2 \right)^2 \frac{M_1 M_2}{M_2^2 - M_1^2}\,
\frac{m_2^2 - m_1^2}{m_1 m_2}\,
c_{12}^2 s_{12}^2 \sin{\Delta}.
\label{ImH}
\ea
Equations (\ref{product})--(\ref{ImH}) are identical
with those of the $\mathbbm{Z}_2$ model,
derived in~\cite{leptogenesis}.
In order to compute the baryon-to-photon ratio of the Universe,
$\eta_B$,
one must~\cite{leptogenesis} multiply $\mathcal{Q}$ by
(i) $M_1 / \left( 10^{11}\, \mathrm{GeV} \right)$,
(ii) a numerical factor of order $10^{-9}$,
(iii) a function of $M_2 / M_1$,
and (iv) $\left( \ln K_1 \right)^{-3/5}$,
where $K_1 \,\propto\, H_{11} / M_1$.
(All these factors are given and explained in~\cite{leptogenesis},
together with references to the original papers.)
One may then compute $\eta_B$ as a function of the input.

Most crucial is the behaviour of $\eta_B$ as a function of $m_1$
when $m_2^2 - m_1^2 = \Delta m^2_\odot$ is kept fixed.
One finds that $\eta_B$ grows with $m_1$,
finding a maximum for $m_1 \sim 4 \times 10^{-3}\, \mbox{eV}$,
afterwards decreasing rapidly for a larger $m_1$.
Now,
the present model---contrary to what happened in the model
treated in~\cite{leptogenesis},
wherein $m_1$ was free---has $m_3 = 0$ and,
therefore,
$m_1 \simeq \sqrt{\Delta m^2_\mathrm{atm}} \sim 0.05\, \mbox{eV}$.
For such a high value of $m_1$
the baryon-to-photon ratio turns out to be hopelessly small.
Thus,
in the present model,
contrary to what happened in the $\mathbbm{Z}_2$ model~\cite{GLZ2}
worked out in~\cite{leptogenesis},
leptogenesis is not a viable option for explaining
the baryon asymmetry of the Universe.

\section{Conclusions}

In this letter we have discussed
an extension of the lepton sector of the Standard Model
with two right-handed neutrino singlets and the seesaw mechanism.
The model,
which was originally proposed in~\cite{softLbar},
is based on the lepton number $\bar L = L_e - L_\mu - L_\tau$. 
Zeros in the $2 \times 3$ neutrino Dirac mass matrix
are enforced by $\bar L$ invariance,
and as a consequence the model features the predictions $\theta_{13} = 0$
and a hierarchical neutrino mass spectrum with $m_3 = 0$.\footnote{These
predictions are common with other models
based on $\bar L$ invariance~\cite{other}.}
The lepton number $\bar L$ is softly broken
in the $2 \times 2$ Majorana mass matrix $M_R$ 
of the right-handed neutrino singlets,
by the two entries $R$ and $S$ in~(\ref{MR})
which would be zero in the case of exact $\bar L$ invariance. 
One obtains $\Delta m^2_\odot \neq 0$
and $\theta_{12} \neq 45^\circ$ from that soft breaking.
However,
$\theta_{12} \sim 33^\circ$ requires the soft breaking to be `strong',
which means that $R$ and $S$
are of the same order of magnitude as the element $M$ in $M_R$
which is allowed by $\bar L$ invariance.
Thus the model discussed here has the property that in $M_R$
there is no trace of $\bar L$ invariance,
whereas the form of the Dirac mass matrix
is completely determined by that invariance.

We have argued that,
for models with one Higgs doublet like the present one,
the configuration $m_3 = 0$ together with $\theta_{13} = 0$
is stable under the renormalization-group evolution.

A further prediction of our model
is the range for the effective mass in neutrinoless $\beta\beta$ decay,
in particular the lower bound given by~(\ref{mbbbound});
the order of magnitude of that effective mass
is the square root of the atmospheric mass-squared difference.

Since there is only one $CP$-violating phase in our model,
we have also considered the possibility of thermal leptogenesis;
however,
it turns out that this mechanism is unable to generate
a realistic baryon asymmetry of the Universe.
This is because in our model the neutrino mass $m_1$ is too large,
due to the inverted mass hierarchy.

In summary,
we have shown by way of a very economical example
that---contrary to claims in the literature---models
based on the lepton number $L_e - L_\mu - L_\tau$
are not necessarily incompatible
with the solar mixing angle being significantly smaller than $45^\circ$.

\vskip 1 cm
\noindent \textbf{Acknowledgements}:
The work of L.L.\ has been supported
by the Portuguese \textit{Funda\c c\~ao para a Ci\^encia e a Tecnologia}
under the project U777--Plurianual.

\end{document}